\documentclass[11pt]{article}
\usepackage{moriond_dpb,epsfig}

\def\be{\begin{equation}}
\def\ee{\end{equation}}
\def\bea{\begin{eqnarray}}
\def\eea{\end{eqnarray}}

\def\vet{\vec E_T}
\def\met{\not\!\!\vet}
\def\gev{~\mathrm{GeV}}
\def\gevc2{~\mathrm{GeV}/\mathrm{c}^2}

\def\D0{D\O}  
\def\d0{D\O}

\begin{document}
\vspace*{4cm}
\title{SEARCHES FOR A HIGH MASS HIGGS BOSON AT THE TEVATRON}

\author{ D. Benjamin \\(for CDF and \d0 Collaborations) }

\address{Duke University,\\Department of Physics Durham, NC USA}
%\\

\maketitle

\abstracts{
 Recent results obtained by the CDF and \d0 collaborations are presented here. These Tevatron Higgs 
searches look for a Standard Model (SM) Higgs boson decaying into W-boson pairs, with the W-bosons decaying into electron-neutrino or 
muon neutrino final states. In the mass range of $135 \gevc2 $ to $200 \gevc2 $, the SM Higgs decays 
prominently into W-boson pairs.  The presented results are based on an integrated luminosity that ranges from 
$3.0$ to $4.2~\mathrm{fb}^{-1}$. No significant excess over expected background is observed and the $95\%$ CL limits are set 
for a Standard Model (SM) Higgs boson for different mass hypotheses ranging from $100 \gevc2$ to 
$200 \gevc2$. The combined Tevatron results exclude SM Higgs boson mass of $160 < m_{H} < 170~\gevc2$. 
}

%\section{Introduction}

 \section{CDF and \d0 experiments at Tevatron collider}
 
  The Tevatron is a $p\bar p$ collider with an center of mass energy of $\sqrt{s}=1.96~\mathrm{TeV}$ operating at the Fermilab.
During the {\it Run II} data-taking period, which started in March 2002, it has delivered more than 
$6~\mathrm{fb}^{-1}$ to the CDF and \d0 experiments.
Results presented in this proceeding are based on data samples with integrated luminosity ranging from $3.0$ to $4.2\mathrm{fb}^{-1}$ depending on the analysis.
 
CDF and \d0 experiments are multi-purpose cylindrical detectors. The azimuthal angle $\phi$ and the pseudo-rapidity $\eta\equiv -ln{[tan(\theta/2)]}$, 
where $\theta$ is the polar angle defined relative to the proton beam axis are the basis of the coordinate system. 
 
CDF consists tracking detectors consisting of a silicon micro-strip detector array surrounded by a 
cylindrical drift chamber in a $1.4~\mathrm{T}$ axial magnetic field. Outside of the tracking chambers, the energies of 
electrons and jets are measured with segmented sampling calorimeters; the outer-most detectors are layers of steel instrumented with planar 
drift chambers and scintillators used for muon identification.
\d0 detector has a central tracking system which consists of a silicon micro-strip tracker and a central fiber tracker, 
both located within a $2~\mathrm{T}$ axial magentic field; a liquid-argon/uranium calorimenter. Finally muons are identified by detectors comprising o
f layers of tracking detectors and scintilators surroundign $1.8~\mathrm{T}$ toroid magnets. Both detectors are described in detail elsewhere \cite{CDFDetector}~\cite{DZeroDetector}.

The transverse energy, $E_T$, is defined to be $E\sin\theta$, where $E$ is the energy associated with a calorimeter tower or cluster, and $p_T$ 
is the component of the track momentum transverse to the beam line. The missing transverse energy vector $\met$, is defined as the opposite of the 
vector sum of the $E_T$ of all calorimenter towers,corrected for the $p_T$ of leptons candidates which do not deposit all of their energy in the calorimeter.

\section{Higgs searches at the Tevatron}

 According to SM, Higgs boson has total cross-section $\times$ branching ratio of about $0.47~\mathrm{pb}$ for a Higgs mass of $165 \gevc2$ \footnote{Higgs boson properties 
depend on its mass, which is not predicted inside the SM framework; here and whenever not explicitly mentioned in the following, numbers refer as an example to an Higgs boson of 
mass $m_{H}=165 \gevc2$.}. At the Tevatron there are four main production mechanisms. The dominant mechanism is via gluon fusion and fermionic loop ($\sim 79\%$), 
followed by Higgs producted in association with a $W$ boson ($\sim 9\%$) or a $Z$ boson ($\sim 5\%$) and finally by vector-boson fusion (VBF) ($\sim 7\%$).

 SM Higgs decay depends on its mass $m_H$. For $m_H > 135\gevc2$ the main decay channel is a W-boson pair, and searches of this final state at Tevatron are commonly defined as 
{\it High Mass Higgs Searches}. The high mass search region is $ 120 \gevc2 ~<~m_H  200~< \gevc2$.

 CDF and \d0 look for both W's decaying leptonically, selecting events with two electrons or two muons or one electron and one muon for a total branching 
ratio of the $WW$ pair of $\sim 6\%$, including the leptonic decay of the $\tau$. The more frequent decay of W-bosons into hadrons is not used due to high 
level of QCD background. 
The di-lepton final state offer a clean signature with manageable backgrounds at hadron colliders; high-$p_T$ electrons and muons are easily selected in the trigger system.

 Drell-Yan (DY) events with two electrons or muons  are the largest background. DY can be suppressed by requiring large missing energy in the event as there are no neutrinos produced
in a DY event. After Drell-Yan suppression the main backgrounds are: heavy di-boson production ($WW$, $WZ$, $ZZ$); $t\bar t$ (particularly for events containing jets); instrumental backgrounds 
arising from $W/Z+\gamma$ or $jets$ events where the photon or the jet is misidentified as a lepton. 
 Most of these processes are modelled using PYTHIA Monte Carlo and a GEANT-based simulation of the detectors. 
An important exception is the $WW$ background: CDF models NLO effects using a pure NLO simulation, namely MC@NLO \cite{mcatnlo}
, while \d0 uses Sherpa to model the $p_T$ of the $WW$ system \cite{sherpastudy}. 
These predictions are then normalized to NNLO cross section calculations for $WH, ZH, t\bar t$ processes, NLO for $VBF$, $WW$, $WZ$, $ZZ$, $W\gamma$. 
 The gluon fusion signal process has been simulated using the most recent calculation available \cite{FlorianGrazziniggH}
which uses the recent MSTW2008 parton density functions (pdf) set \cite{MSTW2008}.
 Data-driven methods are used in order to model the constribution of the instrumental background.

\section{Analysis description}

 To select signal events both collaborations require two high-$p_T$ opposite sign isolated leptons. 
In order to increase acceptance to Higgs events, dedicated analyses also look for final state containing two leptons with the same charge; they will be briefly discussed in sections~\ref{D0SSana} and \ref{CDFSSana}. 
CDF requires the first (second) lepton to have $p_T$ greater than $20$ ($10$)$\gev$, while \d0 asks both leptons to have $p_T$ ($E_T$) greater than $10$ ($15$)$\gev$ for muons (electrons). A significant transverse missing energy $\met$ is then required to reduce DY. The invariant mass of the lepton pair must be greater than $16(15)\gev$ at CDF(D\O) in order to suppress heavy flavor decays and DY events.

 The decay kinematics are used to distinguish the Higgs signal from the much larger backgrounds using multivarient techniques. For example, the opening angle between the final state 
leptons is the strongest discriminant. The spin correlation of the spin 0 Higgs Boson decaying into two spin-1 W bosons imply that the leptons tend to go in the same 
direction. In SM $WW$ events there is no spin correlation between the leptons from the $W$ bosons and the leptons tend to decay back to back.   An Artificial Neural Network (NN) is 
trained to separate signal from background for each different Higgs mass hypothesis. 

\subsection{ \d0 opposite sign analysis}

The \d0 collaboration separates the analyses
% in three different channels, 
depending on the flavor of the final state leptons: $ee$, $e\mu$, $\mu\mu$, with a collected integrated luminosity of, respectively,  $4.2$, $4.2$ and $3.0~\mathrm{fb}^{-1}$.
 Input to the NNs can be classified in three different types: lepton specific variables (e.g. $p_T$ of the leptons), kinematic properties of the whole event (e.g. $\met$) or 
angular variables (e.g. $\Delta\phi(leptons)$).

Table \ref{tab:yieldspreselD0} 
shows the number of expected signal $(m_H = 165 \gevc2)$ and background events including the number of candidates observed in the data.

 No significant excess over predicted background is observed in \d0 data. 
The 95\% CL upper limits on the production cross section   
(using a modified frequentist approach ($CL_s$) \cite{CDFDZeroCombinationPaper})of the SM Higgs boson ($m_H=165\gevc2$) is measured to be $1.7\cdot\sigma^H_{SM}$ 
 with an expected sensitivity of $1.3\cdot\sigma^H_{SM}$. \cite{D0HWWPubNote}.

\begin{table}[th]
\caption{Number of expected and observed events for \d0 $H\rightarrow WW$ opposite-sign analyses. statistical uncertainties only.\label{tab:yieldspreselD0}}
\vspace{0.4cm}
\begin{center}
\begin{tabular}{|l|c|c|c|}
\multicolumn{4}{c}{ \D0 Run II Preliminary $\int \mathcal{L} = 3.0-4.2 \; \rm{fb}^{-1}$, $m_H=165 \gevc2$ } \\
\hline
 Channel & Signal & Background & Data \\
\hline
$ee$ &     $6.13 \pm 0.01$ & $ 332 \pm 15 $ & $336$ \\
\hline
$e\mu$ &   $12.2 \pm 0.1$  & $ 337 \pm 10 $ & $329$ \\
\hline
$\mu\mu$ & $4.85 \pm 0.01$ & $ 4325 \pm 24$ & $4084$ \\
\hline
\end{tabular}
\end{center}
\end{table}

\begin{figure}
\begin{center}
\psfig{figure=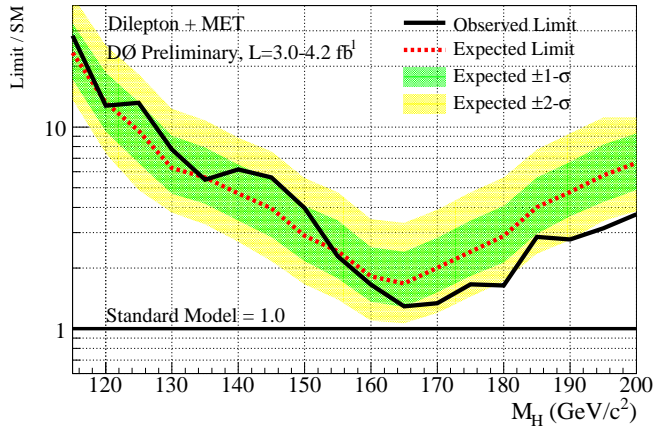,height=2.3in}
\end{center}
\caption{Observed and expected (median, for the background-only hypothesis) $95\%$ CL limits on the ratios to the SM cross section, as functions of the Higgs boson mass for the combined \d0 analyses.
\label{fig:D0Limits}}
\end{figure}

\subsection{ \D0 Same sign analyses}
\label{D0SSana}

 The \D0 collaboration has looked for the process $WH\rightarrow WWW\rightarrow \ell^{\pm}\ell^{\pm}$  where $\ell = (\mathrm{e}~\mathrm{or}~\mu)$ using
$1.0~\mathrm{fb}^{-1}$ of data.. 
 The main backgrounds are instrumental backgorunds coming from charge mis-identification or jets faking a lepton signature. 
For a SM Higgs mass of $m_H ~=~ 160 \gevc2$, the measured limit  is  $24 \cdot\sigma^H_{SM}$ with an expected sensitivity of 
$18 \cdot\sigma^H_{SM}$ \cite{D0SSPubNote}.  This measurement is not included in the Tevatron combination discussed in section \ref{sec:tevcomb}

\subsection{CDF opposite sign analysis}

 CDF colaboration separate the $3.6~\mathrm{fb}^{-1}$ selected sample depending on jet multiplicity, optimizing different NNs for each sample. 
Jets are requested to have $E_T>15\gev, \vert \eta\vert < 2.5$. Input variables to the NNs are analogous to those used by \d0. The specific choice of variables is 
dependent on the jet multiplicty. CDF has chosen to perform
the analysis in this manner as it was found to be the most sensitive; the neural nets are more sensitive to signal identification and background rejection.
Events with no jets have signal contribution only from gluon fusion and the dominant background is $WW$ production. Events with one jet have additional 
signal contribution from associate Higgs production and VBF and $WW$ still remain the main source of background. Finally, signal events with two or 
more jets are dominated by $WH$,$ZH$ and $VBF$ production mechanisms and the dominant background comes from $t\bar t$.

Table \ref{tab:yieldspreselCDF} 
shows the number of expected signal $(m_H = 160 \gevc2)$ and background events including the number of candidates observed in the data.

 The CDF collaboration combines the results from the three opposite sign analyses.
 No significant excess over predicted background is observed in CDF data.
The 95\% CL upper limits on the production cross section   
(using a Bayesian technique \cite{CDFDZeroCombinationPaper}) of the SM Higgs boson ($m_H=165\gevc2$) is measured to be $1.48 \cdot\sigma^H_{SM}$
 with an expected sensitivity of $1.49\cdot\sigma^H_{SM}$. \cite{CDFHWWPubNote}.

%CDF details
%CDF results
\begin{table}[th]
\caption{Number of expected and observed events for CDF\ $H\rightarrow WW$ opposite-sign analyses. (Jet $E_T>15\gev, \vert \eta\vert < 2.5$ ) \label{tab:yieldspreselCDF}}
\vspace{0.4cm}
\begin{center}
\begin{tabular}{|l|c|c|c|}
\multicolumn{4}{c}{ CDF Run II Preliminary $\int \mathcal{L} = 3.6 \; \rm{fb}^{-1}$, $M_H=160\gevc2$} \\
\multicolumn{4}{c}{ opposite sign leptons ($\mathrm{e or} \mu$) }\\
\hline
 Channel & Signal & Background & Data \\
\hline
0 Jets                &  $9.47 \pm 1.46$  & $ 637 \pm 69 $ & $654$ \\
\hline
1 Jet                 &  $5.98 \pm 0.77$  & $ 278 \pm 35 $ & $262$ \\
\hline
$\ge \mathrm{2}$ Jets & $4.53 \pm 0.52$   & $ 173 \pm 23$ & $169$ \\
\hline
\end{tabular}
\end{center}
\end{table}

\subsection{CDF Same sign analysis}
\label{CDFSSana}

 In order to increase the acceptance to Higgs events, the CDF collaborations performs a searche with two same charge final state leptons. The main signal contribution comes from $WH\rightarrow WWW\rightarrow l^{\pm}l^{\pm}+X$, where one of the leptons comes from the W boson produced in association with the Higgs. 
In this search the main backgrounds are instrumental backgorunds coming from charge mis-identification or jets faking a lepton signature. 
CDF uses $3.6~\mathrm{fb}^{-1}$ and an analysis technnique similar to the opposite sign analysis to set $95\%$ CL upper limits on the production of a 
SM Higgs boson ($m_H=165\gevc2$) that are $6.2\cdot\sigma^H_{SM}$ with an expected sensitivity of $7.2\cdot\sigma^H_{SM}$. \cite{CDFHWWPubNote} 

% times the expected SM cross section\cite{D0SSPubNote}.

\subsection{CDF combination (opposite sign and same sign analyses) }

 The CDF collaboration combines the results from the opposite sign and same sign analyses.
 No significant excess over predicted background is observed in CDF data.
The 95\% CL upper limits on the production cross section   
(using a Bayesian technique \cite{CDFDZeroCombinationPaper}) of the SM Higgs boson ($m_H=165\gevc2$) is measured to be $1.5 \cdot\sigma^H_{SM}$
 with an expected sensitivity of $1.3\cdot\sigma^H_{SM}$. \cite{CDFHWWPubNote}.

\begin{figure}
\begin{center}
\psfig{figure=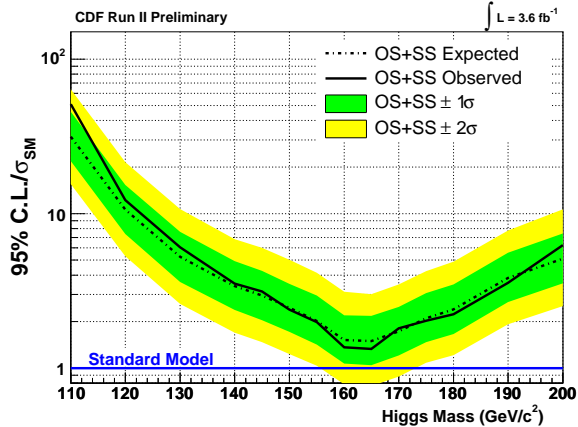,height=2.3in}
\end{center}
\caption{Observed and expected (median, for the background-only hypothesis) $95\%$ CL limits on the ratios to the SM cross section, as functions of the Higgs boson mass for the combined CDF analyses.
\label{fig:CDFLimits}}
\end{figure}

\section{Tevatron combination and results}
\label{sec:tevcomb}

 Results of both collaborations are combined using two different methods: a Bayesian and a modified frequentist technique. \cite{CDFDZeroCombinationPaper}
Both perform a counting experiment for each bin of the final discriminant, including effects from systematics uncertainties. 

 This combination procedure is able to correlate systematics uncertainties among different analyses and experiments. 
Systematics uncertainties are divided in two main categories: rate and shape systematics. 
The rate systematics affect the normalization of the different signal and background contributions; 
these are the most important and are dominated by theoretical uncertainties on signal and background cross sections used to normalize our simulations. 
Shape systematics affect the shape of the output discriminant; an important example is the jet energy scale calibration.

\begin{figure}
%\rule{5cm}{0.2mm}\hfill\rule{4cm}{0.2mm}
%\vskip 1.5cm
%\rule{5cm}{0.2mm}\hfill\rule{4cm}{0.2mm}
\begin{center}
\psfig{figure=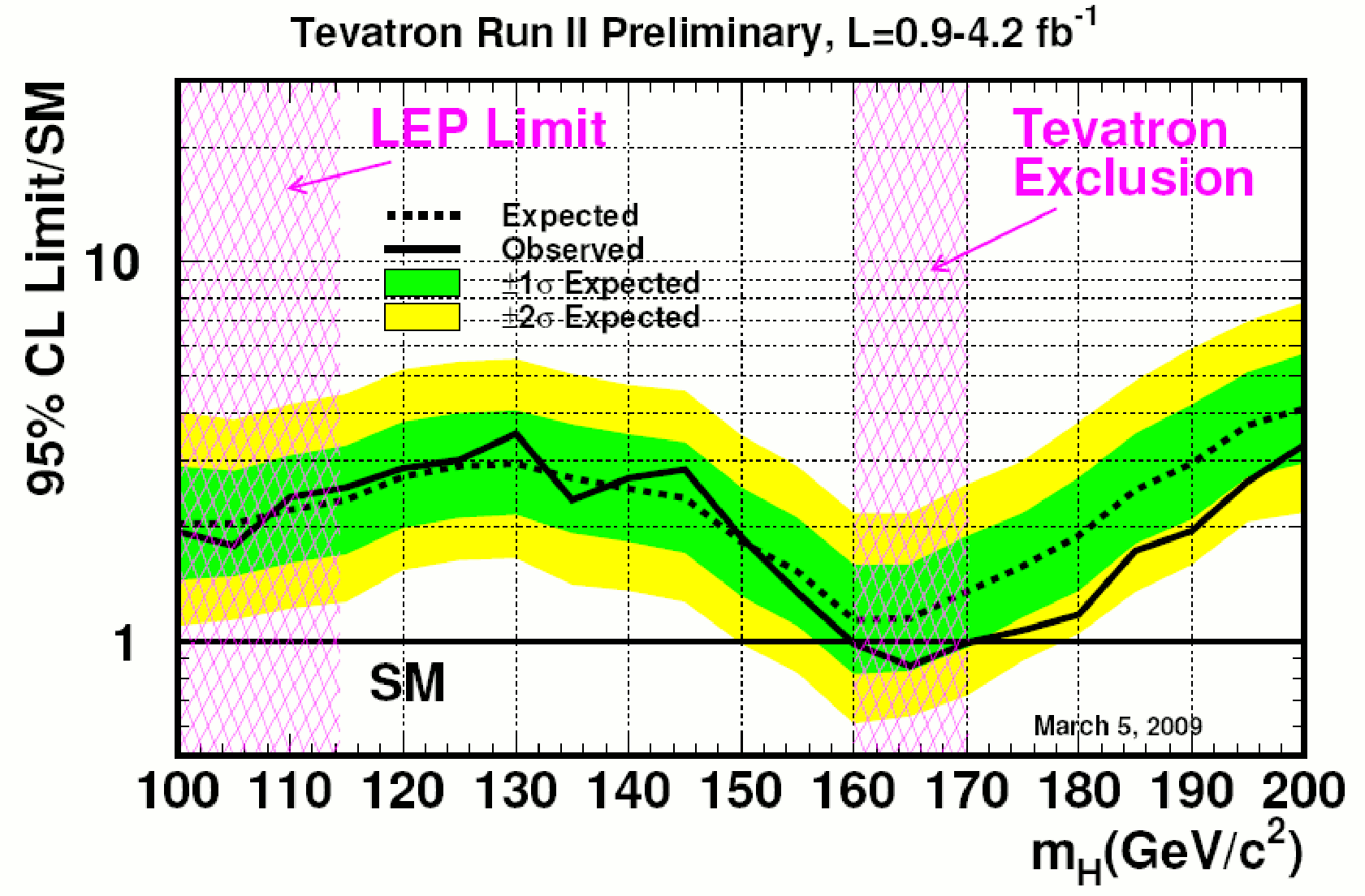,height=2.3in}
\end{center}
\caption{Observed and expected (median, for the background-only hypothesis) $95\%$ CL limits on the ratios to the SM cross section, as functions of the Higgs boson mass for the combined CDF and \d0 analyses.
\label{fig:TevatronLimits}}
\end{figure}

 Figure \ref{fig:TevatronLimits} summarize the combination for Higgs masses between $100$ and $200\gev/c^{2}$.
 The dotted line represent the median, the green and yellow bands one and two sigma spread of the distribution of the expected limits 
from a background-only hypothesis; the solid line is the limit that is set looking at data. Each analysis is performed for different 
Higgs mass hypotheses in $5\gev$ steps.
The combination excludes at 95\% CL a Standrd Model Higgs boson in the $160 < m_H < 170\gev/\mathrm{c}^2$ mass range.
%; the expected limits are $1.1$ and $1.4$ times the expected SM cross section respectively for an Higgs mass of $160$ and $170\gev$.

\begin{table}[th]
\caption{ Tevatron Preliminary (95 \% CL upper limits)$/~\sigma^H_{SM}$ vs $M_H ~(\gevc2)$ Both the Basyesian and $CL_S$ methods results are shown. The more conservative
Bayesian results are quoted.\label{tab:TEVcombined}}
\vspace{0.4cm}
\begin{center}
\begin{tabular}{|c|c|c|c|c|c|c|c|c|c|c|}
\hline
 Basyesian & 155 & {\em 160} & {\em 165} & {\em 170} & 175 & 180 & 185 & 190 & 195 & 200 \\
\hline
Expected & 1.5 & 1.1 & 1.1 & 1.4 & 1.6 & 1.9 & 2.2 & 2.7 & 3.5 & 4.2 \\
\hline
Observed & 1.4 & {\bf 0.99} & {\bf 0.86} & {\bf 0.99} & 1.1 & 1.2 & 1.7 & 2.0 & 2.6 & 3.3 \\
\hline
\hline
 $CL_S$ & 155 & {\em 160} & {\em 165} & {\em 170} & 175 & 180 & 185 & 190 & 195 & 200 \\
\hline
Expected & 1.5 & 1.1 & 1.1 & 1.3 & 1.6 & 1.8 & 2.5 & 3.0 & 3.5 & 3.9 \\
\hline
Observed & 1.3 & {\bf 0.95} & {\bf 0.81} & {\bf 0.92} & 1.1 & 1.3 & 1.9 & 2.0 & 2.8 & 3.3 \\
\hline
\end{tabular}
\end{center}
\end{table}

%Combination methodology
%Main systematics
%Combination result

\section{Conclusions}
 The CDF and \d0 collaborations have continued to improve their sensitivity to low cross section processes;
they now have the sensitivity to observe the SM Higgs Boson. 
The combination of the analysis carried out by the two experiments has led for the first time to the exclusion of a SM Higgs in the $160-170\gev$ mass range. More data is available to be analyzed and more will be collected in the next years allowing an exclusion by each experiment and widening the combined exclusion region.

\section*{Acknowledgments}
I would like to thank organizers of this conference for a wonderful and enlightening conference. Simone Pagan Griso who helped me with the preperation of these proceedings.
The CDF and \d0 collaborations are funded by a whole bunch of government organizations and would like to acknowledge and thank them.

\end{document}